# Disentangling the origin of degradation in perovskite solar cells via optical imaging and Bayesian inference.

*Akash Dasgupta[1], Robert D. J. Oliver[2,1*], Manuel Kober-Czerny[1], Charlie H. G. Nicholls[2,3], Xueli Cao[4,5], Yen-Hung Lin[4,5], Alexandra J. Ramadan[3,1*], Henry J. Snaith[1*]*

[1] Department of Physics, University of Oxford, Clarendon Laboratory, Parks Road, Oxford, OX1 3PU, UK
[2] School of Chemical, Materials and Biological Engineering, University of Sheffield, Sir Robert Hadfield Building, Mappin Street, Sheffield, S1 3JD, UK
[3] School of Mathematical and Physical Sciences, University of Sheffield, Hicks Building, Hounsfield Road. Sheffield, S3 7RH, UK
[4] Department of Electronic and Computer Engineering, The Hong Kong University of Science and Technology, Clear Water Bay, Kowloon, Hong Kong SAR
[5] State Key Laboratory of Displays and Opto-Electronics Technologies, The Hong Kong University of Science and Technology, Clear Water Bay, Kowloon, Hong Kong SAR

* Corresponding authors. E-mails: henry.snaith@physics.ox.ac.uk; a.ramadan@sheffield.ac.uk; robert.oliver@sheffield.ac.uk

**Abstract**

Machine learning and computational inference, coupled with experimental data, promise to significantly accelerate our rate of learning in most scientific disciplines. In this study, we develop tools that connect microscopic observations to macroscopic device behaviour, a capability that is essential for accelerating the design of durable energy materials. To this end, we introduce a novel approach that integrates photoluminescence imaging with drift diffusion simulations to understand operation and degradation in fully fabricated perovskite solar cells. By employing Bayesian inference, we generate 'inferred maps' of parameters that govern recombination processes present in devices. We track these parameter maps while the devices are aged (70 °C, full spectrum sunlight) to analyse their temporal evolution during degradation. Notably, our approach allows us to distinguish between degradation occurring at the hole or electron transporting layer interface, or within the bulk. Our analysis reveals pronounced spatially non-uniform degradation, with significant macroscopic heterogeneity observed in the optoelectronic parameter maps. We pinpoint the greatest degradation observed in specific regions to stem from the perovskite/transport layer interfaces. Finally, we demonstrate that an amino-silane molecular passivation treatment suppresses this degradation, highlighting its specific role in enhancing device stability. Our approach offers valuable insights for future device fabrication and is a clear exemplification of how advanced Bayesian inference can significantly increase the value of experimental data.

# 1 Introduction

More than a decade after their initial demonstration as active materials for solar cells,[1,2], metal halide perovskites have evolved from a laboratory curiosity to a focal point of industrial research and development[3]. The high power conversion efficiencies (PCEs) of 27 % for single junction devices and 34.9% for perovskite-on-Si tandem photovoltaics[4] have generated significant interest within both academic and industrial circles, with stakeholders arguing tha t this technology could play a significant role in addressing global energy challenges[5,6].

Despite impressive progress, achieving the stability and reliability required to make this technology competitive with silicon photovoltaic modules remains a key challenge[7,8]. The performance of perovskite solar cells (PSCs) degrades under a variety of stressors which may be present in normal operation. While extrinsic stressors like $O_2$, $H_2O$, or UV exposure can be mitigated through encapsulation and protective coatings, intrinsic stressors, such as above-band-gap light, voltage bias, and heat, are unavoidable during device operation[9]. The ionic nature of perovskites, distinct from conventional "covalent" PV semiconductors, contributes further complexity to their long-term stability. Redox reactions involving iodide, common in solar-relevant perovskites, can increase ionic activity over time, creating iodide vacancies which can modulate non-radiative recombination, which ultimately limits solar cell voltage[10,11]. Such vacancies also facilitate halide ion mobility, which has been shown to screen the electric field in PSCs, hindering charge extraction, a degradation mechanism that worsens with aging[12]. While the outlook seems positive given current trends in addressing this issue[13], continued effort and further understanding of the mechanisms driving degradation are necessary to make the stability aspect of PSCs as competitive as their efficiencies[14–16].

Importantly, degradation in perovskite solar cell devices extends beyond the bulk perovskite material. Indeed, stability improvements from engineering the perovskite/transport layer interface highlight that complex degradation pathways involving both bulk and interface processes are present in full device stacks[17–21]. Disentangling bulk and interfacial degradation remains challenging, underscoring the need for more device-level studies and a broader range of characterisation methodologies. Moreover, degradation rarely proceeds uniformly across the device; instead, it tends to evolve heterogeneously, necessitating spatially resolved techniques for effective characterisation[22,23]. Particularly, imaging techniques capable of probing full devices at micron to centimetre scales have emerged as powerful tools[23–25]. Such techniques have revealed significant heterogeneity in perovskite solar cells[23,26–31], and have confirmed that degradation often nucleates at pre-existing defects rather than occurring uniformly[22,32–36].

In our previous work, we demonstrated how photoluminescence (PL) imaging of fully fabricated cells under accelerated aging conditions can reveal device-scale degradation dynamics[37]. Spatially resolved quasi-Fermi level splitting (QFLS) and "charge collection quality" ($Q_{col}$) maps[23] revealed that localised regions of poor luminescence and extraction gradually expand across the device. This degradation was significantly suppressed using an amino-silane passivation strategy, yielding relatively high-performing PSCs (~23 %). Through a variety of additional techniques, such as 'Time-of-Flight Secondary Ion Mass Spectrometry' and ab initio simulations, we concluded that the silane molecule would coordinate to the top surface of the perovskite active layer, and thus infer that the observed suppression of degradation results from this interaction. However, while powerful, the measured QFLS and $Q_{col}$ maps alone were not sufficient to establish a direct correlation between the observed degradation (which was occurring spatially heterogeneously on the device) and bulk or surface degradation

Herein, we present a new approach which overcomes previous limitations, enabling the spatial disentanglement of distinct degradation pathways in fully fabricated perovskite solar cells using intensity-dependent photoluminescence imaging. Our previous interpretations of PL imaging were limited by the complexity of the physical processes. Here, we combine the imaging with drift-diffusion simulations (choosing to use a simulation package that explicitly account for the influence of mobile ionic species), modelling how the QFLS evolves with illumination intensity.

Crucially, we apply Bayesian inference methods[37] to extract the most probable values of key physical parameters from the experimental PL data. By performing this inference on a per-pixel basis, we generate spatially resolved 'inferred parameter maps'. Importantly, the inclusion of mobile ions in our drift-diffusion model enables us to distinguish recombination at the electron and hole transport layer interfaces independently- a separation that typically requires dual-side measurements and is not feasible in standard "mono-facial" solar cell architectures. This unique capability allows us, for the first time, to quantify the relative contributions of bulk and interfacial degradation processes in a spatially resolved manner across the entire device area.

Our results uncover clear signatures of heterogeneous surface degradation that originates in large-scale defect regions and expands over time, as well as millimetre-scale domains with distinct bulk degradation characteristics. In contrast, devices with amino-silane passivation exhibit remarkable spatial-uniformity, with extracted parameters remaining stable throughout aging. This methodology provides an effective means for understanding how degradation manifests in operational perovskite devices: not just in aggregate, but as a spatially and physically resolved phenomenon on relevant length scales.

## 2 Method and theory

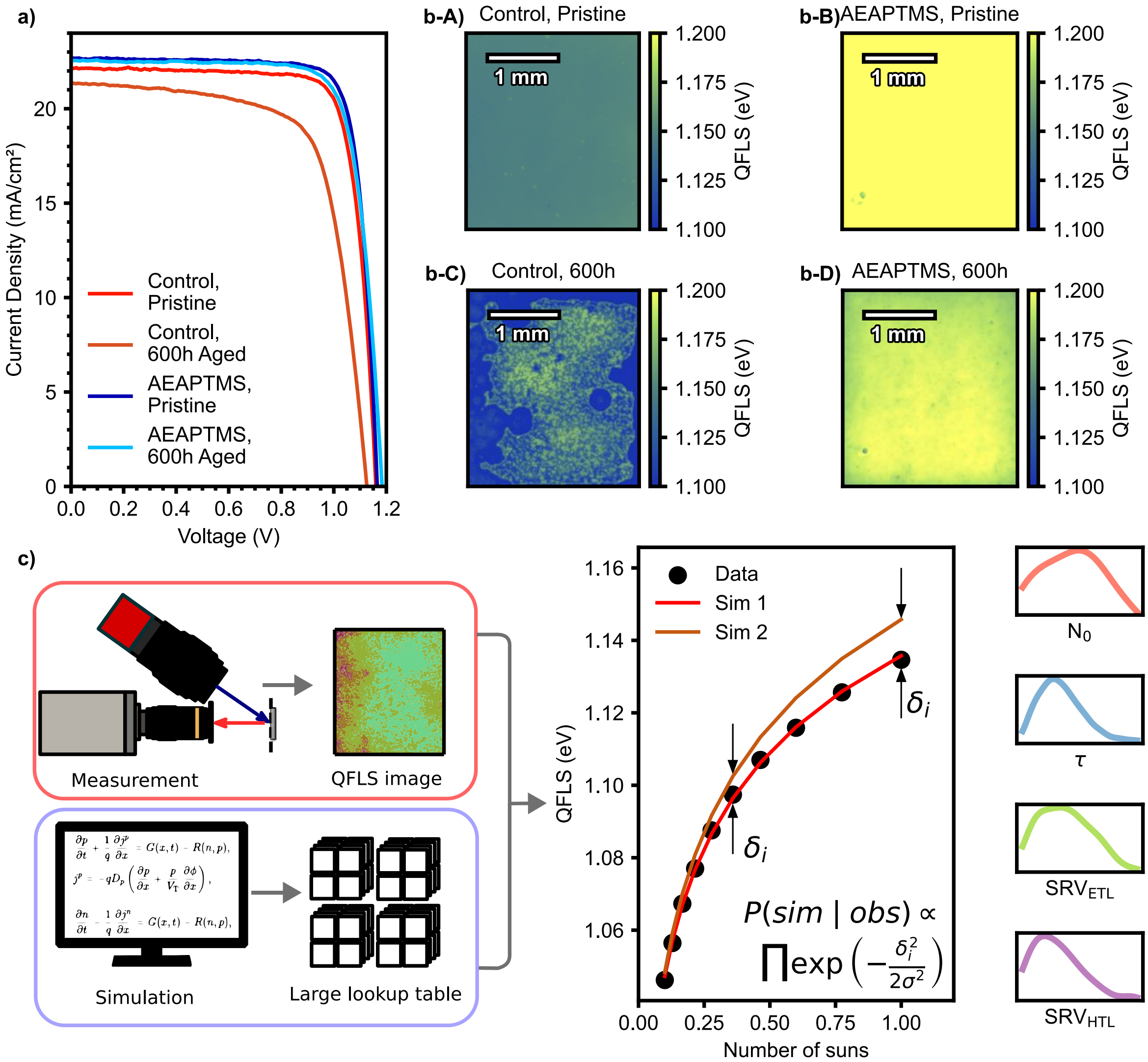

**Figure 1**: (a) Current density–voltage (*JV*) curves for control and AEAPTMS-treated perovskite solar cells with the architecture ITO/Me-4PACz/$Cs_{0.13}FA_{0.87}Pb(I_{0.9}Br_{0.1})_3/C_{60}$/BCP/Cr/Au. The devices were subjected to accelerated aging under simulated full-spectrum sunlight at 70°C in ambient air and open-circuit conditions (Closely following ISOS-L-2 standard). (b) Spatially resolved quasi-Fermi level splitting (QFLS) images of *(i)* control (pre-aging), *(ii)* AEAPTMS-treated (pre-aging), *(iii)* control (600 h aged), and (iv) AEAPTMS-treated (600 h aged) devices, at 1-sun equivalent illumination. Scale bars represent 1 mm. Aged control devices show notable QFLS loss at the edges, indicating spatially inhomogeneous degradation. (c) Schematic overview of the modelling workflow: QFLS maps are interpreted using drift-diffusion simulations, with Bayesian inference applied to extract the most probable underlying physical parameters.

Our goal in this work is to extract spatially resolved physical parameters from intensity-dependent PL measurements. As we mentioned, this work was motivated by a previously published dataset on the degradation of AEAPTMS-treated and control perovskite solar cells under accelerated aging[37]; herein we will apply our new analysis to the data from that publication. Because the mechanism behind the passivation is thoroughly explored in the previous work, by doing this advanced analysis, we can compare the outputs of our approach to a pre-established understanding. In doing so, we will also provide a more quantitative, spatially resolved picture of how charge carrier recombination evolves during aging, allowing spatially localized degradation features to be examined directly and assigned to either surface- or bulk-dominated effects.

Figure 1 presents the core experimental dataset that underpins our analysis. The solar cells used in this study featured a $Cs_{0.13}FA_{0.87}Pb(I_{0.9}Br_{0.1})_3$ perovskite active layer with a photovoltaic band gap of approximately 1.6 eV. The device stack consisted of: indium tin oxide (ITO)/Me-4PACz ([4-(3,6-dimethyl-9H-carbazol-9-yl)butyl]phosphonic acid) self-assembled monolayer (SAM) hole transport layer/perovskite/$C_{60}$ electron transport layer/bathocuproine (BCP)/chromium/gold. Further fabrication and encapsulation details can be found in our previous report[41]. Panel (a) shows current density-voltage (*JV)* curves for both pristine "control" and AEAPTMS-treated devices and the same devices following over 600 h of accelerated aging (sealed, under full-spectrum simulated sunlight at 70 °C under open-circuit conditions in ambient air, corresponding to International Summit on Organic Solar Cells Stability(ISOS)-L-2 standard)[42–44]. The control device exhibits significant degradation, while the AEAPTMS-treated device retains most of its initial performance.

Figure 1b presents spatially resolved QFLS maps acquired under 1-sun equivalent illumination over the same aging period. These maps reveal a spatially localised nature of the degradation, revealing the emergence of localised "dark regions" in the control cell, particularly near the device edges. Although the AEAPTMS treatment clearly mitigates these effects, the QFLS maps alone are unable to determine whether the molecule primarily stabilises the bulk perovskite or passivates one or both interfaces. In our previous work, it was determined that the AEAPTMS aids stability primarily by providing effective surface defect passivation. Density functional theory (DFT) and ab initio molecular dynamics (AIMD) simulations revealed that the AEAPTMS molecule binds strongly to the perovskite surface, eliminating the effects of surface vacancy defects, which we inferred to inhibit degradation pathways. In addition to simulations, time-of-flight secondary ion mass spectrometry (TOF-SIMS) showed that AEAPTMS molecules remained predominantly within the surface region, with minimised bulk penetration. Hence, we inferred that the 'dark regions' are primarily surface degradation, but our analysis of the QFLS fell short of making this determination in and of itself.

To overcome this shortcoming here, we employ Bayesian inference: a method of statistically inferring the values of parameters, based on a model (either analytical or numeric), from a set of measurements. In PV, this has been used successfully to determine underlying material parameters from *JV* measurements[45–47], and in the perovskite community it has been gaining traction recently in its application on analysing time-resolved photoluminescence measurements[42]. The utility in the method (compared to traditional fitting algorithms) is in its handling of degeneracy, that is, when multiple combinations of input parameters in a model can produce outputs consistent with the measurements. Bayesian inference addresses this challenge by sampling a broad set of parameters (such as carrier lifetime, ion density, etc.), evaluating the model for each instance, and calculating a probability value (described below) indicating the likelihood that the sampled parameters are the true parameters underlying the measurements. Doing this for models evaluated over a range of parameters (e.g., QFLS at different bulk carrier lifetimes) allows us to build a 'posterior probability distribution' of that parameter.

In this work, we follow the approach of Brandt et al. [48], by modelling the recombination physics of the solar cell using a drift-diffusion simulator, and running a large set of simulations spanning different combinations of physical parameters. Specifically, we used the 'IonMonger' package[42], which is able to carry out simulations where mobile ion vacancies are coupled to the charge carrier dynamics. By generating a large set of simulations, we construct a database (or look-up table) that maps parameter permutations to corresponding simulated intensity-dependent QFLS values. This 'look up table' can then be combined with our measured QFLS data using Bayesian inference, on a per-pixel basis, allowing us to infer posterior probability distributions for the underlying physical parameters for the first time in a spatially resolved map (Figure 1c). We chose to use this look-up table approach as in Brandt et al. [45] in favour of the 'Markov chain Monte Carlo' (MCMC) sampling methods we used previously while modelling time-resolved PL[49–51], because the spatially-resolved nature of our dataset involves a very large number of simulations: one for each pixel, with ~1000 pixels per 300 μm x 300 μm image. While MCMC inference benefits from the freedom to explode the full parameter space (as opposed to our approach, which works within a pre-computed parameter space domain), doing this for each pixel would be computationally infeasible. By front-loading the computation into the creation of the look up table, we eliminate the need for further simulations during inference, making this approach both efficient and scalable: A first of its kind applied to perovskites.

To determine the QFLS and charge collection quality ($Q_{\mathrm{col}}$) maps, we used a home-built setup comprising a blue 450 nm LED illumination source and a scientific-complimentary metal-oxide-semiconductor (sCMOS) camera, and a source meter to electrically contact the sample, as in our previous works[23,37,49]. We acquired PL images of the samples under various illumination intensities. We estimated PL quantum yield (PLQY) maps by dividing the PL maps imaged through a 515 nm long-pass filter, by an image of a white diffuse reflector reference directly imaged with no filter, at each illumination intensity, taking the latter as a proxy for 100% incident photons (accounting for the wavelength response of the detector and optical components). This is converted into QFLS maps by the relationship:

$$QFLS = QFLS_{\mathrm{rad}} + kT\ln(PLQY)\,, \quad (1)$$

where $k$ is the Boltzmann constant, $T$ absolute temperature, and $\mathrm{QFLS}_{\mathrm{rad}}$ the maximum QFLS possible given the material bandgap and illumination intensity. Full details can be found in Supplementary Information section 1 and Figure S7, and associated calibrations and error estimates can be found in our previous report where we use this setup for the first time[23].

In IonMonger, we simulate a cell at open-circuit, with varying illumination intensity of 0.01 sun to 1 sun equivalent in logarithmic steps. To replicate the experimental procedure, each illumination intensity within the simulation is run for a simulated time of 30 seconds to allow for ionic rearrangement. Full details can be found in Supplementary information section 6. Finally, we combine experiment and simulation by considering the probability a particular simulation matches the observation, given by

$$P(sim|obs) \propto \exp\left(\sum_{i=1}^{n} -\frac{(QFLS_{\mathrm{obs}} - QFLS_{\mathrm{sim}})_i^2}{2\sigma^2}\right), \tag{2}$$

Where $i$ denotes the different intensities measured between 1 and $n$, and $\sigma$ the error associated with the measurement and $QFLS_{obs}$ and $QFLS_{sim}$ is the observed and simulated QFLS respectively. This is derived from Bayes theorem. To obtain the posterior distribution for an individual parameter $m$ (e.g., bulk carrier lifetime), we marginalise over all other parameters by summing the probabilities of all simulations that share the same value of $m$:

$$P(m|obs) = \sum_{sim \in S(m)} P(sim|obs), \tag{3}$$

where $S(m)$ denotes the set of simulations with parameter $m$ fixed and all other parameters varied. This yields the marginal posterior probability distribution for $m$. This workflow is illustrated in Figure 1c, and more information is available in Supplementary Information Section 2. By evaluating this probability across our full simulation look-up table, we construct pixel-wise posterior distributions over the input parameters, allowing us to quantitatively infer the most likely physical parameters underlying spatial variations in recombination behaviour. These inferred parameters can then be used to interpret the mechanisms driving the observed variations.

In creating the look up table, it is unfeasible to vary every possible input parameter in a drift diffusion simulation, since the time required to run simulations increases as a power of the number of parameters (and we have over 25 possible input parameters). To select the most physically meaningful and computationally tractable subset, we follow the formalism developed by Courtier et al.[51]. In this framework, the QFLS is governed not only by recombination processes but also by the electric potential landscape induced by mobile ions. Redistribution of these ions leads to the formation of electrostatic double layers at the perovskite/transport layer interfaces, causing the potential to drop sharply at these locations rather than smoothly across the absorber (Fig. S4)[53–55]. These potential drops are inherently asymmetric at the ETL and HTL interfaces, and because the QFLS response depends on the location of recombination relative to these fields (Fig. S5), the model can distinguish between surface recombination occurring at the ETL versus the HTL side. Crucially, the inclusion of the parameter of ion vacancy density ($N_0$) directly controls the strength and asymmetry of the interfacial fields and thus modifies how the device responds to light intensity. This means $N_0$ not only governs ionic behaviour but also enhances the model's sensitivity to the spatial origin of interfacial recombination.

Based on this, we chose to vary four key parameters: the bulk lifetime $\tau$, the surface recombination velocities at the ETL and HTL interfaces ($SRV_{\mathrm{ETL}}$, $SRV_{\mathrm{HTL}}$), and the ion vacancy density $N_0$, with the other values being fixed by expected values in the literature. This set allows us to capture the dominant recombination pathways and ionic effects influencing the QFLS. The sensitivity of QFLS and ideality factor to these specific parameters is further supported by independent literature[56,57]. For the simulations, we selected a range of 25 logarithmically spaced values. Full details and a parameter list may be found in Supplementary information section 6, in Tables S1, S2.

This approach provides, for each pixel on our QFLS image, a probability distribution for our varied parameters. To interpret these probability distributions in a more accessible form, we compute the *expected value* of each parameter. In statistical terms, the expected value, $E(m)$, is the mean of a probability distribution, defined as

$$E(m) = \int P(m|obs) \cdot m \, dm, \tag{4}$$

where $m$ is a stand-in for specific simulated parameter values and $P(m|obs)$ is the posterior probability density. This operation collapses the full distribution into a single representative value, allowing us to meaningfully compare parameter trends across the device area and over the aging period. Since our QFLS measurements are spatially resolved, we apply this calculation on a per-pixel basis, producing *inferred parameter maps* that visualise the spatial variation in the underlying recombination and ion-related physics. A more detailed description of this implementation is provided in Supplementary Information Section 3.

# 3 Results and discussion

## 3.1 Elucidating full solar cell degradation

### 3.1.1 Less severe degradation

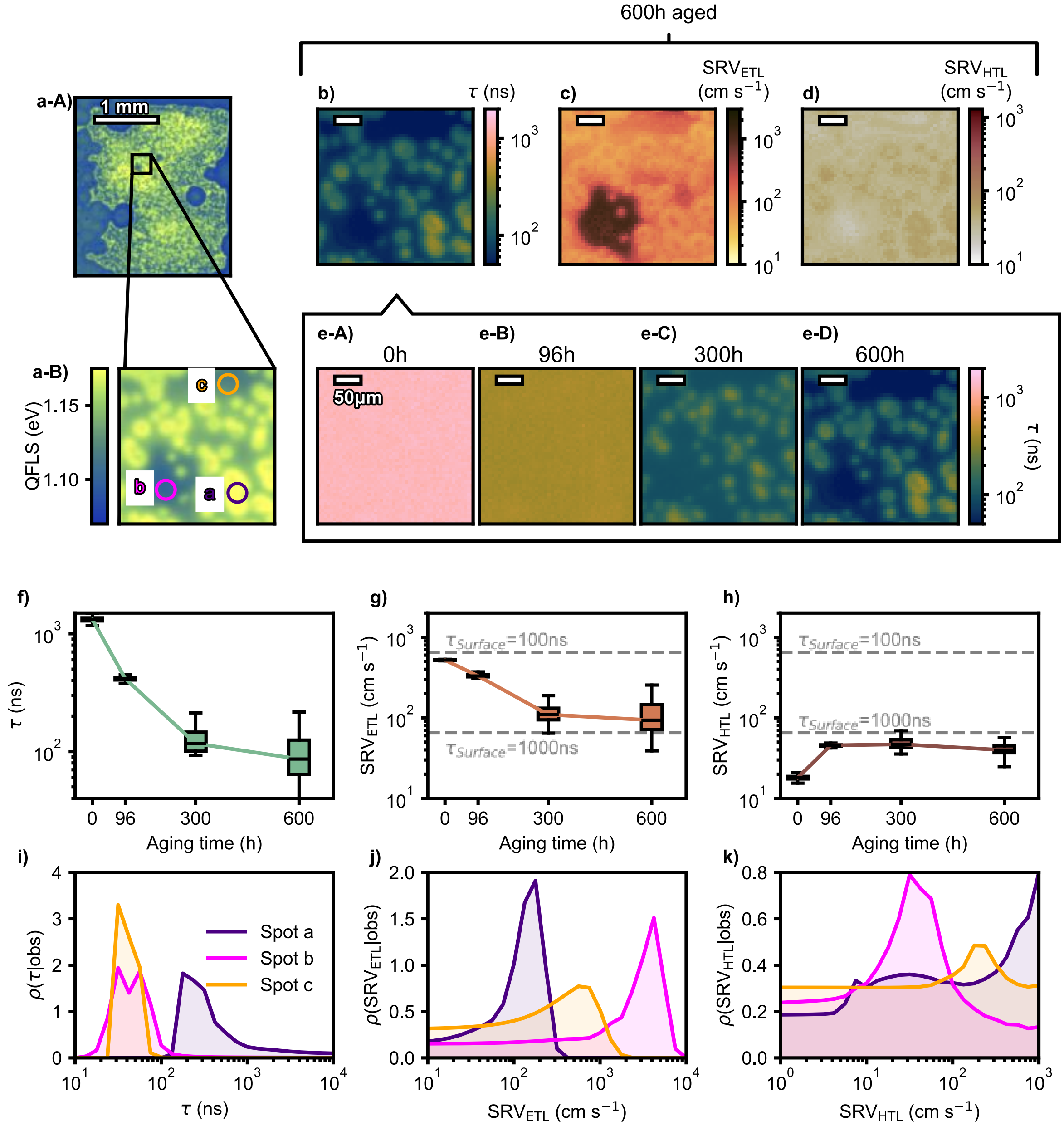

**Figure 2:** Spatially-resolved Bayesian inference of recombination parameters in a 'less degraded' region of the control perovskite solar cell after 600 h of aging. **(a**) *(A)* QFLS image of the full control device area (0.5 mm x 0.5 mm) at 600 h of aging, showing heterogeneous degradation. A region with relatively high QFLS, indicative of less degradation, is highlighted. (B) Zoomed-in view of the highlighted region (~0.3 mm x 0.3 mm), revealing local heterogeneity and bright clusters. Three representative pixels (a, b, c) are marked for detailed analysis. **(b–d)** Inferred parameter maps for the zoomed-in region at 600 h aging, showing *(b)* Expected values for bulk carrier lifetime ($\tau$), *(c)* Expected values for surface recombination velocity (SRV) at the electron transport layer (ETL) interface, and *(d)* Expected values for SRV at the hole transport layer (HTL)

interface. Values correspond to the expected values of the Bayesian posterior distributions per pixel. **(e)** Evolution of the inferred bulk lifetime ($\tau$) map at aging times of 0 h, 96 h, 300 h, and 600 h, shown in panels *(A-D)*. **(f–g)** Boxplots of the distributions of inferred values across the maps for each aging timepoint: *(f)* bulk lifetime, *(g)* SRV at the ETL and HTL interfaces. Boxplots represent the median and interquartile range of the *expected values* from the posterior distributions, aggregated across all pixels in the mapped region; each boxplot summarises the distribution of per-pixel expected value parameter estimates, not the underlying full posterior distributions themselves. **(i–k)** Full posterior probability density distributions at points a, b, and c (from panel a-B), plotted on the same axes for comparison. (i) Bulk lifetime, (j) $SRV_{\mathrm{ETL}}$, and (k) $SRV_{\mathrm{HTL}}$. These distributions illustrate the uncertainty and variability in parameter inference at the pixel level.

We begin our analysis by investigating degradation within the control device. As seen in Fig. 2a-A, the QFLS map after 600 h aging can be separated into two distinct regions: The areas near the edges where the QFLS is uniformly low, and areas where the average QFLS remains higher, but composed of discrete luminous clusters. For our analysis, we selected a small (~0.3 mm x 0.3 mm) region placed inside the 'higher QFLS' area, as indicated in Figure 2a. In Figure 2a-B), we can see the luminous clusters more clearly. These features are regularly distributed across the region in an isotropic manner, rather than stemming from one specific spot and growing outwards.

Applying our Bayesian approach to this region, we can generate inferred parameter maps , shown in Figure 2b. We stress that the values depicted here are the 'expected value' of each parameter, the calculated average value weighted to the probability distribution as described in Equation (4, and detailed fully in Supplementary information section 4. We will discuss the trends derived from these 'expected values' first, before analysing more deeply the underlying posterior distributions.

In the inferred parameter maps shown in Figure 2b-d, it is clear that the maps of $\tau$, the bulk charge-carrier lifetime, most closely resemble the raw QFLS map, suggesting local variations in this parameter are the dominant factor governing the appearance of the brighter QFLS clusters. While the maps for the SRV on both interfaces also have spatial structure, the most pronounced contrast between the bright and dark regions are seen in the maps of the expected value of $\tau$. Additionally, the expected value of $\tau$ in the 'darker' regions (interstitial to the clusters) is <100 ns, meaning the bulk recombination in these regions is dominant over any surface recombination (SRVs all < 1000 cm $s^{-1,}$ corresponding to a surface lifetime of > 100 ns in a first order approximation).

Tracking the $\tau$ expected value maps at different points through the accelerated aging process (Figure 2e), there is an initial uniform reduction of lifetimes. However, with continued aging, the bright clustered regions maintain longer lifetimes, potentially even showing slight "photo-brightening"[57], while the interstitial areas exhibit a sharp decline. In Figure 2f, where the median value for the expected value of $\tau$ drops significantly over the course of the accelerated aging, with an increasing interquartile range representing the increasing heterogeneity between the clustered and interstitial regions. This hence points to a spatially heterogeneous, yet isotropic, degradation of the perovskite bulk material itself. Some regions retain high lifetimes, while others degrade into highly nonradiative-recombination-dominated regions, but these regions themselves are uniformly distributed across the image area rather than originating and remaining localised around any specific point.

In contrast, the median expected value for $SRV$ at the ETL interface surprisingly slightly decreased over the aging period, i.e., the surface lifetime increases (Figure 2g), which aligns with a slight increase in the median QFLS as the device is aged (Fig. S4). Several plausible mechanisms could explain this reduction in $SRV$. This could result from interfacial chemical processes, such as oxidation-induced passivation [60], despite encapsulation. Alternatively, mobile ions may redistribute halide vacancies, passivating surface defects, or changes in ETL carrier density, e.g., Fermi level shifts not accounted for in the model, could lead to an apparent reduction in $SRV$[10–12,32,58–60]. In order to verify that this trend extracted from the device data is meaningful, we performed an independent, orthogonal measurement (time-correlated single photon counting, TCSPC). Using TCSPC, we extract bulk lifetimes and surface recombination velocities at the perovskite/ETL interface following well established protocols by ourselves and others (Figures S8-15). Full details are provided in the Supplementary Information. These measurements allowed us to confirm the surprising trend inferred from our model, where the $SRV_{ETL}$ reduces following aging Figure 2g. The materials parameters extracted from the TCSPC measurements are shown in Table S3. The values extracted from these fittings show very good agreement with then median expected values derived from our Bayesian method here: the lifetime drops from 700ns to 240ns, while the $SRV_{ETL}$ decreases from ~1000 cm $s^{-1}$ to 500 cm $s^{-1}$. Not only are the trends of declining lifetimes and declining $SRV_{ETL}$ reproduced, the actual values are close to the median behaviour inferred by our Bayesian method.

Interestingly, while the median value of $E(SRV_{\mathrm{ETL}})$ decreases, we observe a localised region of elevated expected $SRV_{ETL}$ values in the spatially resolved map (Figure 2c), in the same region marked 'b' in Figure 2a-B. This also coincides with low collection quality (Supplementary Figure S2), indicating localised surface degradation may be emerging in this region. More broadly, this example demonstrates the value of spatially resolved inference: rather than relying solely on median or bulk trends, our approach reveals local variations and emerging degradation patterns that would otherwise be obscured in spatially averaged measurements.

Finally, we can access the full potential of using a Bayesian approach by analysing the full posterior distributions at selected spatial locations, allowing us to draw a more nuanced view of the uncertainty in the trends previosuly described. We will examine the full posterior distributions at three representative spatial locations, marked as points (a), (b), and (c) in Figure 2a-B. These were chosen to represent distinct regimes: (a) lies inside a bright cluster, (b) is in a darker area where the $SRV_{\mathrm{ETL}}$ is inferred to be higher, and (c) an interstitial area where the $SRV_{\mathrm{ETL}}$ was predicted to be lower. Additionally, full 'corner plots' for these points are provided in the supplementary text (Fig S12-S15), which provides 2D visualisation of the higher dimension parameter corelations.

The posterior plots of $\tau$ and $SRV_{\mathrm{ETL}}$ (at 600 h aging time) are relatively well-defined, exhibiting sharp, distinct peaks, which indicate strong confidence in the extracted expected values. Moreover, the distributions for points (a), (b), and (c) are clearly separated, confirming that the spatial heterogeneity discussed in the maps Figure 2a and b reflects meaningful variation in the underlying recombination parameters, since the expected values are derived from these distinct distributions with minimum overlap.

The posterior distributions for $SRV_{\mathrm{HTL}}$ however requires greater discussion. The full posterior distributions for $SRV_{\mathrm{HTL}}$ (Figure 2k) are notably broader than those of the other parameters, which may initially imply that the expected values extracted are untrustworthy. Specifically, while the distribution in 'spot b' seems to have a clear peak, 'spot c' has a much broader distribution, while spot 'a' has a broad peak at approximately 12 cm $s^{-1}$ , but also a high probability at higher values (>100). In the case of spot b, we have a high level of confidence that the $SRV_{HTL}$ is relatively low, since there is a clear peak in the posterior distribution. Our previous analysis stands: The limiting factor for the QFLS here is the bulk lifetimes. In spot c, while the distribution is broader, there is still a clear peak at a relatively low value. Additionally, if we study the corner plots in figure S22, specifically looking at the cross corelation between $SRV_{HTL}$ and $SRV_{ETL}$ we can see that when represented on a 2D plot, the peak is much clearer. Hence, in this case, while we are much less confident in the specific 'expected value' derived from the posterior, we can judge it lies within the order of magnitude of the posterior peak (~100 cm $s^{-1}$). The inferred bulk lifetime in this spot was less than 100ns, meaning that once again, our previous analysis still stands: The QFLS is limited by the bulk, rather than the surface.

In 'spot a', the analysis is less certain due to the strong 'edge' present in the posterior at higher values for $SRV_{HTL}$, implying the possibility that the low QFLS observed in this spot may be caused by HTL degradation. However, we can discard this possibility by looking at the cross-corelations between the $SRV_{HTL}$ and the other parameters on the corner plot in Figure S20. Particularly, looking at the cross corelation between $SRV_{ETL}$ and $SRV_{HTL}$ reveals that the predicted high $SRV_{HTL}$ occur only for lower values of $SRV_{ETL}$, ie, the $SRV_{HTL}$ is only predicted to be high when the $SRV_{ETL}$ is low (<100 $cm^{-2}$). However, looking at the other cross-corelations (particularly the $\tau$ vs $SRV_{ETL}$ corelation plot), it is clear the $SRV_{ETL}$ is predicted to have a higher value in this spot (>100 $cm^{-2}$), with a very clear peak in the 2D corelation plot. Hence, we can recognise the high probability tail for $SRV_{HTL}$ as an artefact of representing the 4-dimentional probability distribution into 1 dimension and discard the possibility of HTL degradation on this spot. This example highlights the utility of the Bayesian framework: not only does it enable spatial mapping of expected parameter values, but it also provides access to the underlying probability distributions, allowing us to critically assess both parameter identifiability and physical plausibility.

While we are also able to generate maps of ion density, $N_0$, from our analysis, we choose to omit these. The posterior distributions obtained for this parameter were too wide for us to reasonably assign an expected value to this parameter. We plot the averaged posterior distribution for $N_0$ (ie, the distribution obtained by averaging the posterior distributions of all pixels in the image slice) in Supplementary figure S16. There it is clear that the posteriors for the reference samples on average do not show clear peaks, such that the value obtained by taking the statistical expected value would be meaningless for comparison. Hence, from the intensity-dependent QFLS measurements alone, we are unable to resolve the increasing ion density with aging which is often reported in the literature[61]. This does not mean that the ion density does not change during aging in this system, only that our measurements are not sensitive to the specific value of this parameter. Inclusion on the ion density into the drift diffusion model is still required to distinguish the $SRV_{ETL}$ from the $SRV_{HTL}$ (discussed fully in Supplementary note 5).

### 3.1.2 More severely degraded region

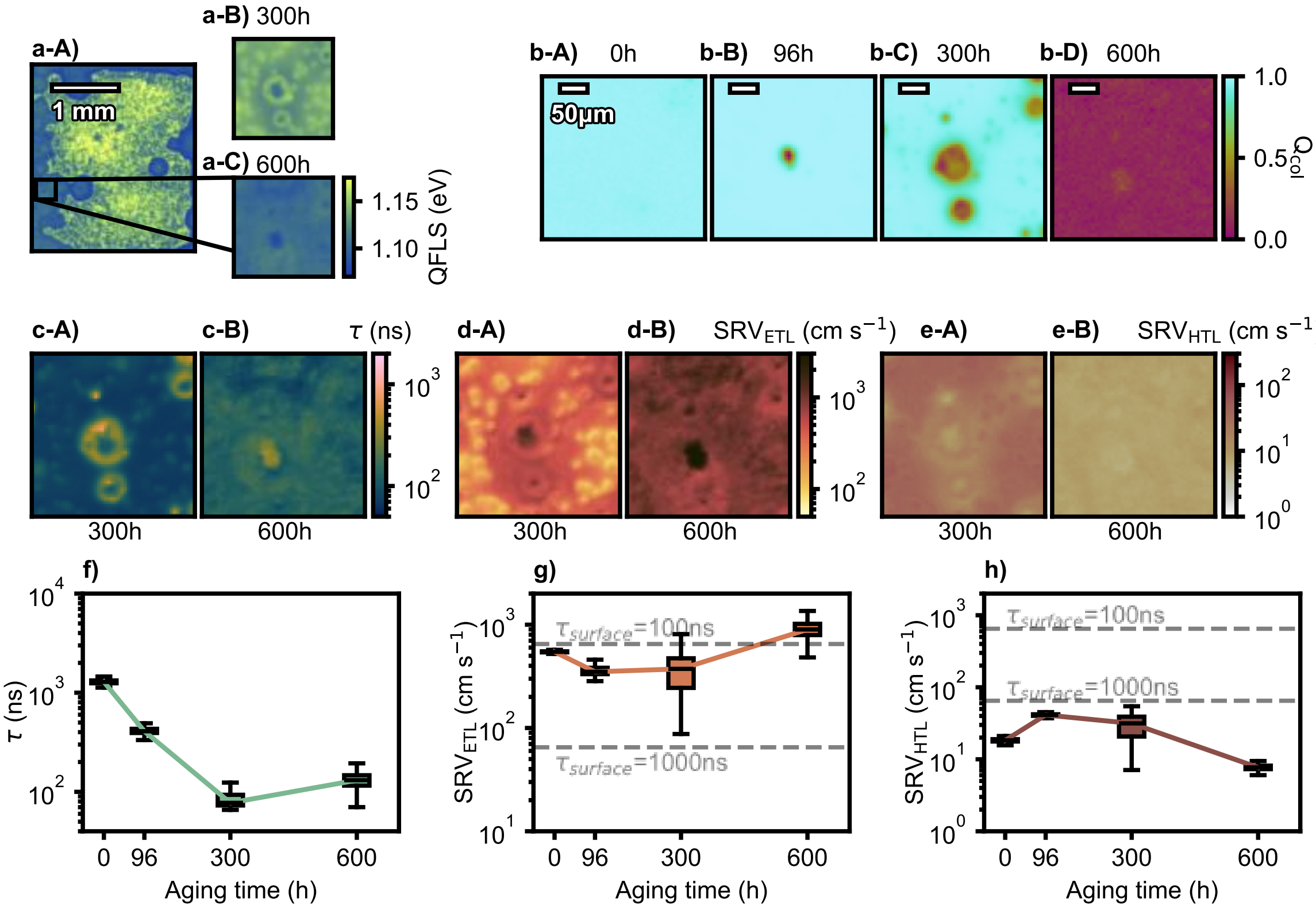

**Figure 3:** Spatially-resolved Bayesian inference of recombination parameters in a more degraded region of the control perovskite solar cell. **(a)** *(A)* QFLS image of the full control device area (0.5 mm × 0.5 mm) at 600 h of aging, same as in previous panel, showing heterogeneous degradation. *(B-C)* Zoomed-in views (~0.3 mm × 0.3 mm) of a lower-QFLS region with more pronounced degradation, shown at *(B)* 300 h and *(C)* 600 h aging. **(b)** Maps of the derived charge collection quality ($Q_{col}$) in the zoomed-in region at aging times of *(A)* 0 h, *(B)* 96 h, *(C)* 300 h, and *(D)* 600 h, calculated from PL measured at short-circuit and open-circuit conditions. **(c–e)** Inferred parameter maps in the same region at 300 h and 600 h aging for**: (c)** expected bulk carrier lifetime (d) surface recombination velocity at the ETL interface ($SRV_{ETL}$), and (e) SRV at the HTL interface ($SRV_{HTL}$) at 300 h (A) and 600 h (B) of aging. **(f–h)** Boxplots of the distributions of expected inferred parameter values across the maps for each aging timepoint: (f) bulk lifetime (τ), (g) $SRV_{ETL}$, and (h) $SRV_{HTL}$. Boxplots show the median and interquartile range of the expected values from the Bayesian posterior per pixel, aggregated over the full mapped region.

We now turn to a more severely degraded region of the same control device, focusing on the area highlighted in Figure 3a. As we discuss in our previous report[62], degradation in the QFLS in this region appears to originate from discrete defective points visible at earlier aging times (Figure 3a-B) which expand over time into contiguous areas of low QFLS (Figure 3a-C). This is clearer in the zoomed-out map of Figure 3a-A, which shows that these low-QFLS regions form approximately circular patterns, corresponding to early-stage defective sites.

Here, we apply our Bayesian inference approach to determine if the degradation observed in this region is a bulk or a surface effect: i.e., is the difference in the degradation driven by localised degradation of the perovskite absorber, or by increased recombination at interfaces? Some evidence for the latter can be found in the charge collection quality ($Q_{col}$) maps, a metric linked to carrier extraction efficiency under short-circuit conditions. The $Q_{col}$ maps show the same circular patterns and outward growth (Figure 3b) consistent with a loss in charge extraction and pointing toward interface-related degradation.

In Fig 3c-B, we show the inferred parameter map for $\tau$. Unlike the 'less degraded' region on the sample (Figure 2b), the expected bulk lifetime at 600h aging is relatively homogeneous, in the 100–300 ns range. Moreover, the early darker (lower QFLS) point visible at 300 h (Figure 3c-A), actually has a slightly *higher* expected value of $\tau$. This indicates that low QFLS in this region at 600h aging is not primarily driven by bulk recombination losses. In contrast, the maps of expected value of $SRV_{\mathrm{ETL}}$ show elevated values in the 'darker point' at 300 h (Figure 3d-A), and in the entire area at 600 h (Figure 3d-B), reaching values of approximately 1000 cm $s^{-1}$. The boxplots in Figure 3–h show that up to 300 h, the median $\tau$ decreases (Fig. 3f) and the median $SRV_{\mathrm{ETL}}$ decreases (Figure 3g), in line with the trend observed in the less degraded region of the sample. However, at 600 h aging, the median $E(SRV_{\mathrm{ETL}})$ increases significantly (Figure 3g).

This indicates that up to 300 h most of the device in this region behaved similarly to the 'less degraded' region, with the exception of the darker spot, where the $SRV_{ETL}$ is higher. Since this spot represents only a few pixels in the 300 h aged map, it does not meaningfully skew the boxplots in Figure 3g. We look more closely at the corner plots, if we select a spatial point inside the 'dark spot' (Figure S23) in comparison to a point inside (Figure S24). For the spot outside, the cross corelation between $\tau$ and $SRV_{ETL}$ is very confined, with some 'tails' at higher $\tau$ / lower $SRV_{ETL}$. Inside the dark spot, the posterior is a little broader, allowing for at lower $\tau$ / higher $SRV_{ETL}$ or higher $\tau$ / lower $SRV_{ETL}$ being a possible explanation for the reduced luminescence, but the 'peak value' still well defined at a high SRV value. The peaks in the 1D projections are also well defined in the 2D projections (save for the $N_0$ cross corelations) , showing our interpretation of the expected value maps are accurate and not skewed by projection artefacts.

By 600h the entire region has clearly deviated from the less degraded region, exhibiting uniformly high $SRV_{ETL}$. Thus, while degradation in this region initially follows the same trends as in the less degraded region, an additional degradation mode emerges, confined to a small "spot" at 300 h but expanding outward by 600 h to encompass a much larger area. There are small shifts to higher values in median $\tau$ between 300 h and 600 h (Figure 3f,h), but these should not be interpreted as signs of improved bulk performance. once ETL-side recombination dominates, the sensitivity of QFLS to other recombination pathways is reduced. This is evident in the averaged 1D projections of the posterior distributions shown in the Supplementary Information: The $\tau$ distributions (Fig. S17) broadens significantly at 600h, shifting the median to higher values.

Taken together, the elevated $SRV_{ETL}$ in the localised dark spot, which grows outward over time, combined with the corroborating $Q_{col}$ data in Fig. 3b, indicates that this additional degradation mode originates from ETL surface/interface recombination. It is initiated in isolated regions and progressively expands, forming larger areas of poor device performance.

## 3.2 Passivated perovskite solar cell degradation

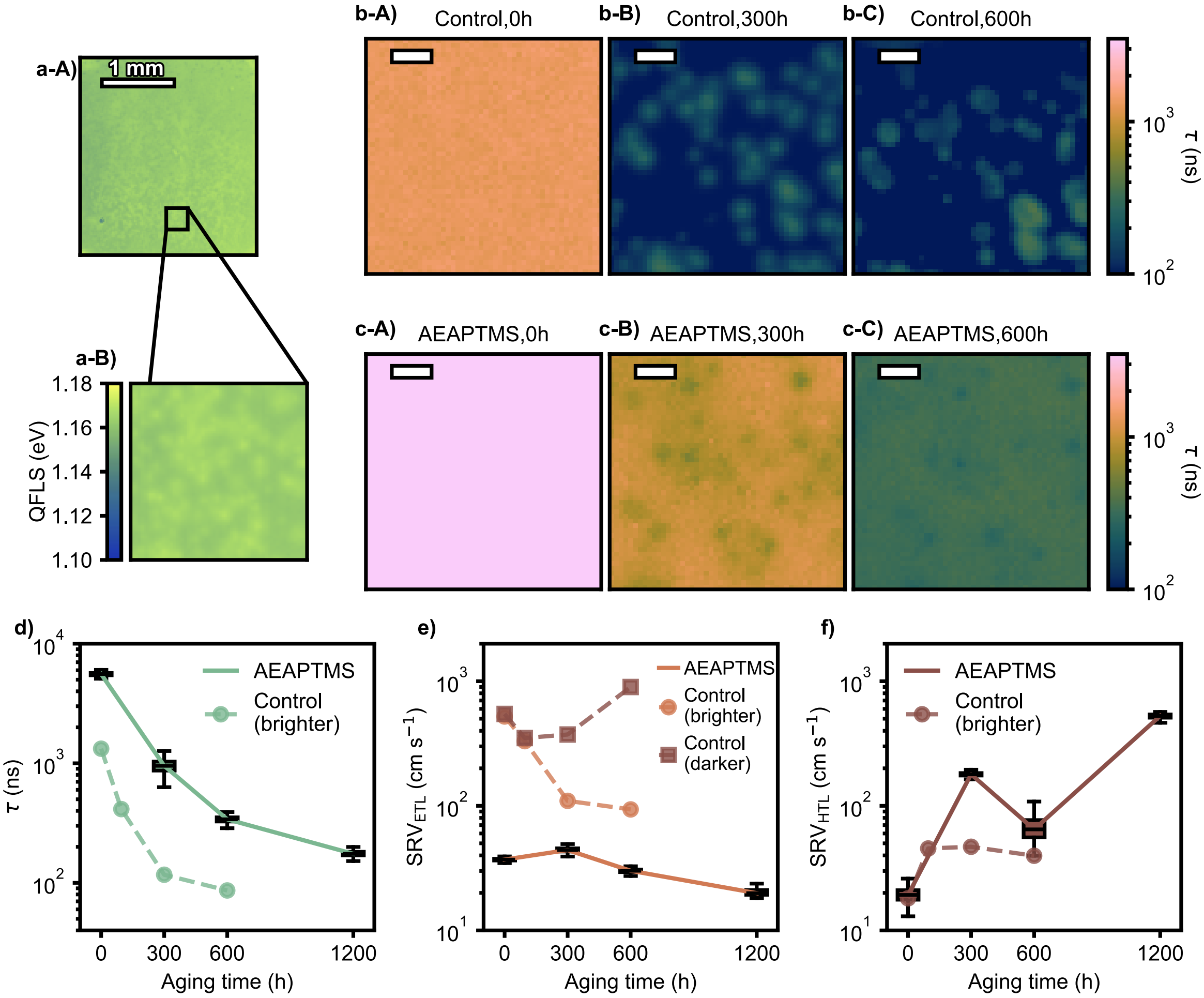

**Fig 4:** Comparison of recombination parameter stability in AEAPTMS-passivated and control devices. (a) (A) QFLS image of the AEAPTMS-passivated device at 600 h aging, showing relatively uniform luminescence across the device area. (B) Zoomed-in view (~0.3 × 0.3 mm) of a representative region selected arbitrarily due to the spatial uniformity of the QFLS image. (b–c) Inferred maps of the expected bulk carrier lifetime for the (b) less degraded region of the control device (from Fig. 2) at (A) 0 h, (B) 300 h, and (C) 600 h aging, and for the (c) AEAPTMS-passivated region at (A) 0 h, (B) 300 h and (C) 600 h aging. All parameter maps use a consistent spatial scale; scale bars = 50 $\mu$m. (d–f) Boxplots of per-pixel expected values from the posterior distributions for each aging timepoint in the AEAPTMS-passivated device, for (d) $\tau$, (e) $SRV_{\mathrm{ETL}}$, and (f) $SRV_{\mathrm{HTL}}$. For comparison, median values from the less degraded control region are shown as overlaid scatter points, connected by dashed lines to visualize temporal evolution. In panel (e), both the less and more degraded control regions are shown in the scatter overlay.

Finally, we analyse results from our passivated device. We focus on a specific region with the same dimensions as studied above to enable easier comparison with the un-passivated device. We note the high degree of homogeneity of the PL images, meaning comments about this region may be taken to apply to the whole device.

To recall, our previous research highlights the effectiveness of the amino-silane passivation strategy in enhancing device stability and efficiency[37,62]. Specifically, we showed via molecular dynamics simulations that this molecule applied between the perovskite and electron transport layer (ETL), exhibits effective binding to halide vacancy sites at the surface. This, taken together with TOF-SIMS measurements which showed the molecule remains localised to the surface suggested its role is to passivate the ETL/perovskite interface[63]. We do acknowledge however that recent work by Huang et.al has also indicated that further chemical reactions may be occurring between the primary amine and formamidinium[64,65], with the in-situ synthesis of new passivating compounds, as observed with other amine passivation routes[63]. Armed with this understanding, we can evaluate the effectiveness of our Bayesian approach in revealing more detailed understanding and certainty in the mechanistic impact of this passivation route. The observations and insights from our previous study were not incorporated into the model here in any way. Hence, the inferred values produced may be considered as *a priori,* because our previous understanding was not used as an input and hence cannot affect the output of our approach.

For the AEAPTMS-treated sample, the QFLS map remained notably uniform compared to the control (Fig 4a). In the control device, the appearance of luminous QFLS 'clusters' is attributed by our analysis to a significant reduction in local bulk lifetime in the surrounding regions, with only isolated high-lifetime regions remaining. In contrast, no such features were observed in the passivated device (Fig 4b,c). In the fresh state, the expected values for $\tau$ is much higher in the AEAPTMS treated sample (Fig 4b-A,c-A). While there is a decline in bulk lifetime with aging, the expected values for $\tau$ remains higher than the control sample throughout the process (Fig 4d), and the decline is spatially uniform, without formation of the clusters as seen in the control sample (Fig 4b-C,c-C). This suggests that, independent of whether it is the dominant factor in device stability, the addition of AEAPTMS *does* have a bulk passivation role; The expected bulk lifetimes are higher, and the formation of heterogenous clusters of 'higher' lifetime and 'lower lifetime' interstitial space is supressed. This would be consistent with the results from Huang[63], where they suggest secondary reaction compounds have a bulk passivation effect.

Next, we can turn out attention to the $SRV_{ETL}$ expected values. We recall catastrophic device failure is ascribed by our analysis to localised points of high $SRV_{ETL}$, which grow outwards to encompass larger areas of the device. Applying our analysis to the AEAPTMS device, it shows from the outset a significantly lower SRV at the ETL interface, remaining under 100 cm $s^{-1}$ (equivalent to a surface lifetime of ~1,000 ns). this is consistent with our ab-initio simulations from our previous work, suggesting that the molecule is effective at coordinating to vacancies at the perovskite/ETL surface. Additionally, the $SRV_{ETL}$ remains largely constant as the device is aged: in contrast to the un-passivated control in both the 'less degraded' region, where it decreases over time (as also confirmed by our TRPL measurements); and in the case of the 'more degraded' region in Fig3 A-B., the $SRV_{ETL}$ increases dramatically after sufficient aging, indicating it is surface degradation which leads to catastrophic device failure. This leads us to conclude, effectively *a priori*, that the dominant role of the AEAPTMS treatment in terms of stability is to passivate the perovskite/ETL interface, stabilising it against degradation. Since this is the degradation pathway that leads to the biggest change in QFLS in the control devices, we can conclude that suppression of this degradation channel specifically is what leads to the excellent device stability observed in the AEAPTMS passivated cells.

Once again, in order to confirm the trends predicted by our Bayesian approach, we additionally look at TCSPC measurements conducted on partially fabricated stacks. Fitting the AEAPTMS 'half stacks'[66–68] (Figures S8-13) with fitting values tabulated in Table S3, we find excellent agreement. The passivation when applied (before aging), yields an increased bulk lifetime (~ 700 ns in control to 1000ns in treated devices) and a reduced $SRV_{ETL}$ (~1000 cm $s^{-1}$ in control to 30 cm $s^{-1}$ in treated), with values very close to what our Bayesian approach predicts. After aging, the $SRV_{ETL}$ measured via TCSPC remains low after aging, while the control, as we mentioned earlier, follows the trends from the 'less degraded' region very well. This additional orthogonal characterization further cements our confidence in the trend predicted by our method. As a macroscopic measurement, TCSPC averages over the excitation spot and therefore lacks spatial resolution. Consequently, the signal is dominated by the most emissive regions and is effectively insensitive to localized degradation.

Taken together, our analysis elucidates a twofold effect of the AEAPTMS treatment. We observe both bulk and surface passivation: There is a clear suppression of the heterogenous bulk degradation seen in less degraded regions of the control sample, consistent with recent studies[66]; and, in addition, the localised degradation on the perovskite ETL interface, characterised by a low $Q_{col}$ and higher expected values of $SRV_{ETL}$, is also supressed, with the $SRV_{ETL}$ of the treated cell remaining constant with aging. Since increasing $SRV_{ETL}$ is the cause of the large, low QFLS regions in the control which we associated with device failure, and since the bulk lifetimes still drop with aging (albeit remaining higher than the control at all times), we can conclude that the preservation of the surface defects is the primary mechanism which leads to the longer device stability offered by this passivation molecule.

# 4 Conclusion

In this work, we present a powerful and generalisable framework that combines spatially resolved luminescence imaging, drift-diffusion simulations, and Bayesian inference to extract and visualise key physical parameters within perovskite solar cells. By applying this method to aged devices, we generate quantitative, pixel-level maps (~10 micron resolution) of critical recombination parameters, including bulk lifetime and surface recombination velocities at both transport layer interfaces, and track their evolution over time. This capability moves beyond traditional, spatially averaged analyses, enabling us to localise and identify heterogeneous degradation mechanisms across the device.

Using this approach, we reveal that degradation in standard perovskite solar cells proceeds heterogeneously across length scales of hundreds of microns to millimetres. In less degraded regions, we observe a pattern of QFLS 'clustering', where bright, high-lifetime islands persist as the surrounding material darkens with aging. Our analysis shows that this behaviour is driven primarily by spatial variations in bulk lifetime, identifying bulk recombination as the dominant degradation pathway in these zones.

Conversely, in regions where degradation is more severe, we observe a significant drop in QFLS accompanied by increasing surface recombination velocity at the perovskite/ETL interface. These degraded regions originate from discrete points that expand outward over time, forming the circular degradation patterns seen in the raw QFLS data. Coupled with a simultaneous decline in charge collection quality, our results strongly support a surface degradation mechanism in these areas, with the ETL interface identified as the primary site of failure.

By applying our method to devices treated with an amino-silane passivation layer, we further demonstrate that this strategy effectively suppresses surface degradation at the ETL interface and mitigates bulk degradation. The treated cells maintain high and uniform lifetimes throughout aging, underscoring the dual protective role of the passivation approach and validating our inference results with prior independent chemical and optical characterization.

This work demonstrates, for the first time, a robust, spatially resolved, and statistically grounded methodology capable of deconvolving the contributions of bulk, ETL-side, and HTL-side recombination across a single macroscopic photovoltaic device. By enabling per-pixel inference of underlying physical processes, this approach offers a new paradigm for understanding how local optoelectronic behaviour drives macroscopic device performance and stability. We anticipate that this framework will enable broader studies across perovskite and other emerging photovoltaic technologies, particularly in identifying failure modes, evaluating new materials, and informing targeted passivation strategies. More broadly, it contributes to the growing movement toward data-rich, model-integrated analysis of device function and degradation, from microscopic structure to macroscopic performance.

## Acknowledgments

The authors acknowledge funding from the Engineering and Physical Sciences Research Council (EPSRC UK) grant number EP/X039285/1. This work was funded by UK Research and Innovation (UKRI) under the UK government's Horizon Europe funding Guarantee [grant number 10054976 of the "Next Generation of Sustainable Perovskite-Silicon Tandem Cells" (NEXUS) project and No. 861985 of the PEROCUBE project]. AD and RDJO were funded by the Greenwood and Penrose scholarships respectively. This work was part funded by the DFG (Deutsche Forschungsgemeinschaft) via the SPP2196 Priority Program (CH 1672/3-1). RDJO acknowledges funding from the Royal Society through award RG\R1\241060. CHGN is grateful to the Overli Foundation Scholarship for funding his Studentship. The authors would like to thank Nicola Courtier for informative discussions, and for creating and maintaining the IonMonger software used in this work. The authors would like to acknowledge the use of the University of Oxford Advanced Research Computing (ARC) facility in carrying out this work[69].

## Competing Interests

H.J.S. is co-founder and chief scientific advisor of Oxford PV Ltd., a company commercializing perovskite PV technology.